\documentclass[fleqn,twoside]{article}
\usepackage{espcrc2}
\usepackage{graphicx}
\usepackage{pstricks}
\usepackage[figuresright]{rotating}
\def \pb        {{\rm \, pb}}
\def \fb        {{\rm \, fb}}
\def \ipb       {{\rm \, pb^{-1}}}
\def \ifb       {{\rm \, fb^{-1}}}

\def \GeV       {{\rm \, GeV}}

\def \GeVcc     {\GeV/c^2}

\def\ga{\mathrel{\raise.3ex\hbox{$>$\kern-.75em\lower1ex\hbox{$\sim$}}}}
\def\la{\mathrel{\raise.3ex\hbox{$<$\kern-.75em\lower1ex\hbox{$\sim$}}}}

\newcommand {\bfell}      {\ell\kern-0.4em
                           \ell\kern-0.4em
                           \ell\kern-0.4em
                           \ell }
\newcommand {\obfell}     {\overline{\ell}\kern-0.4em
                           \overline{\ell}\kern-0.4em
                           \overline{\ell}\kern-0.4em
                           \overline{\ell}}
\newcommand {\bfH}      {\; {\cal H}\kern-0.5em \kern-0.4em
                           {\cal H}\kern-0.5em \kern-0.4em
                           {\cal H}\kern0.1em }
\newcommand {\obfH}     {\; \overline{\cal H}\kern-0.5em \kern-0.4em 
                           \overline{\cal H}\kern-0.5em \kern-0.4em 
                           \overline{\cal H}\kern0.1em }
\def \b             {{\mathrm b}}
\def \t             {{\mathrm t}}
\def \charm         {{\mathrm c}}
\def \d             {{\mathrm d}}
\def \u             {{\mathrm u}}
\def \e             {{\mathrm e}}

\def \h             {{\mathrm h}}
 
\def \f             {{\mathrm f}} 

\def \W             {{\mathrm W}}
\def \Z             {{\mathrm Z}}

\def \P             {{\mathrm P}}

\newcommand {\dM}         {\Delta M}

\newcommand {\sfe}     {{\tilde{\f}}}
\newcommand {\sfL}     {{\tilde{\f}_{\mathrm L}}}
\newcommand {\sfR}     {{\tilde{\f}_{\mathrm R}}}

\newcommand {\sneu}    {{\tilde{\nu}}}

\newcommand {\seR}     {{\mathrm{\tilde{e}_{R}}}}

\newcommand {\st}      {{\mathrm{\tilde{\tau}}}}
\newcommand {\stR}     {{\mathrm{\tilde{\tau}_{R}}}}

\newcommand {\smR}     {{\mathrm{\tilde{\mu}_{R}}}}

\newcommand {\sto}     {{\tilde{\mathrm{t}}}}

\newcommand {\sbot}    {{\tilde{\mathrm{b}}}}

\newcommand {\snu}     {{\tilde{\nu}}}

\newcommand {\neu}     {{\chi}}

\newcommand {\thstop} {\mathrm{\theta_{\tilde{t}}}}
\newcommand {\thsbot} {\mathrm{\theta_{\tilde{b}}}}

\newcommand {\tanb}{\tan\beta}
\newcommand {\ee}    {{\e ^+ \e ^-}}

\newcommand {\pmiss}   {{P\!\!\!\,\!/ }}
\newcommand {\emiss}   {{E\!\!\!\,\!/ }}
\title{SUSY Particles Searches at LEP and Interpretations within the MSSM}
\author{Giacomo Sguazzoni\address[CERN]{European Laboratory for
Particle Physics (CERN), CH-1211 Geneva 23, Switzerland}
}

\begin{document}

\begin{abstract}
Searches for R-parity conserving supersymmetric particles 
have been performed in $\ee$ data collected
by LEP detectors, at centre-of-mass energies up to
$209\GeV$, corresponding to an integrated luminosity of $\sim\!3.1\ifb$. The
results and their interpretation in the context of MSSM
frameworks are briefly reviewed.
\vspace{1pc}
\end{abstract}

\maketitle

\section{INTRODUCTION}

Theories with Supersymmetry (SUSY) are the most promising extensions
of the Standard Model (SM)~\cite{SUSYgeneral}.
The simplest version is the Minimal
Supersymmetric Model (MSSM), which contains the minimal number of
additional particles. The scalar fermions or {\em sfermions}, $\sfL$ and
$\sfR$, are the partners of the left- and right-handed SM fermions and
mix to form the mass eigenstates. The mixing angle $\theta_{\sfe}$ is
so defined that $\sfe\!=\!\sfL\cos\theta_{\sfe}\!+\!\sfR\sin\theta_{\sfe}$ is
the lightest sfermion. In general mixing is relevant for the third
family, while $\sfe\!\equiv\!\sfR$ otherwise.
The SM gauge boson states have fermionic super-partners or {\em
gauginos}. The MSSM higgses are arranged into two doublets and their fermionic
super-partners are the {\em higgsinos}. Neutral 
gauginos and higgsinos mix into four mass eigenstates, the
{\em neutralinos} $\chi$, $\chi_2$, $\chi_3$, $\chi_4$
($M_{\chi_4}\!>\!M_{\chi_3}\!>\!M_{\chi_2}\!>\!M_{\chi}$). The charged gauginos 
and higgsinos mix into two mass eigenstates, the {\em charginos}
$\chi^{\pm}$ and $\chi^{\pm}_{2}$ ($M_{\chi^\pm_2}\!>\!M_{\chi^\pm}$).

The lepton and baryon number conservation translates into the
``R-parity'' conservation. 
The LSP (Lightest Supersymmetric Particle) is stable and must be also
neutral and weakly interacting to fit the cosmological observations.
Within the MSSM the LSP is the lightest neutralino $\chi$ or, less
likely, the sneutrino, $\sneu$. At LEP the sparticles are pair
produced and the decay brings to final states 
containing at least one LSP. 

The success of LEP searches is also due to the impressive
performance of the accelerator that provided $\ee$ collisions at 
centre-of-mass energies between $161$ and $209\GeV$,
and an integrated luminosity of $\sim\!775\ipb$ per experiment. 
The results, based on the full
high-energy data sample, are presented in the form of 95\% C.L.
exclusion domains in the space of the relevant parameters, since no
excess has been observed.
When available, the LEP SUSY Working Group
combinations, based on the outcomes from ALEPH, DELPHI, L3 and OPAL
(ADLO), are reported~\cite{LEPSUSYWG}. 

\section{PRIMARY SIGNALS}
\label{sec:signals}

Except few pathological cases, sparticle pair production leads to the typical
acoplanar particles topology due to missing energy ($\emiss$) and
momentum ($\pmiss$) from escaping LSP's. The energy of the visible
system is related to the mass difference between the sparticle $\widetilde{\P}$
and the LSP ($\dM\!=\!M_{\widetilde{\P}}\!-\!M_\mathrm{LSP}$).
The acoplanar topologies studied cover each type of visible final
state (leptons, hadronic jets, $\gamma$'s).

\subsection{Acoplanar leptons}

\begin{figure}[t]
\includegraphics[width=0.36\textwidth]{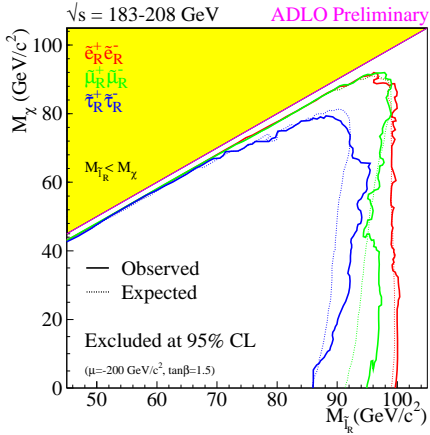} 
\vskip -1cm
\caption{Slepton mass exclusion plot from the LEP SUSY Working Group.}
\label{fig:sleptons}
\end{figure}

The analyses for slepton signals
($\ee\!\to\!\tilde{\ell}^{+}\tilde{\ell}^{-}$, $\tilde{\ell}\!\to\!\ell\neu$) 
search for acoplanar leptons by using
the powerful lepton and tau identification of LEP
detectors~\cite{ALEPH_slept_209,DELPHI_ichep02,L3_ichep02,OPAL_ichep02}. 
The LEP combined cross section upper limits range from $10$ 
to $60\fb$~\cite{LEPSUSYWG}. The resulting mass lower limits are $100\GeVcc$,
$94\GeVcc$ and $86\GeVcc$ for $\seR$, $\smR$ and $\stR$
respectively, valid for $\dM\!>\!10\GeVcc$, as shown in Figure~\ref{fig:sleptons}. 

\subsection{Acoplanar jets}

\begin{figure}[t]
\includegraphics[width=0.34\textwidth]{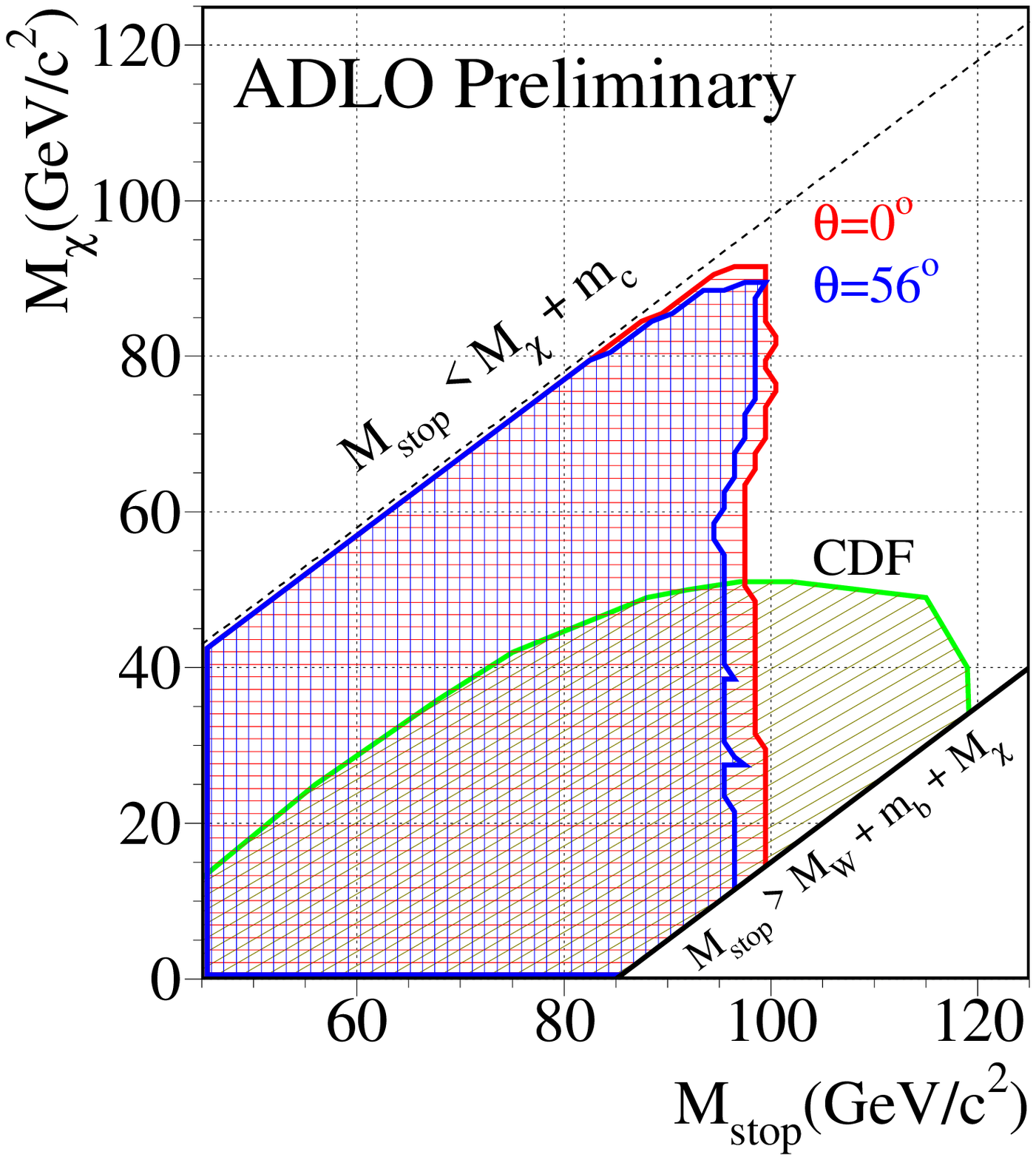} 
\vskip -1cm
\caption{Stop mass exclusion plot from the LEP SUSY Working Group in
case of $\sto\!\to\!\charm\neu$ decay for minimal ($\thstop\!=\!56^\circ$) and maximal
production cross section ($\thstop\!=\!0^\circ$). The CDF result is also shown.} 
\label{fig:squark}
\end{figure}

The production of a squark pair results into an acoplanar
jet topology. These hadronic events can be selected by using event variables and requiring
$\emiss$ and
$\pmiss$~\cite{DELPHI_ichep02,L3_ichep02,ALEPH_squark_209,OPAL_squark_209}.
In case of 
$\ee\!\to\!\sto\bar{\sto}$, $\sto\!\to\!\charm\neu$, the mass lower limit is
$94\GeVcc$ for $\dM\!>\!10\GeVcc$ and any mixing, as visible in
Figure~\ref{fig:squark}. Further specialized selections are used for
other squark processes: b-tagging is effective for
$\ee\!\to\!\sbot\bar{\sbot}$, $\sbot\!\to\!\b\neu$, allowing a limit of
$92\GeVcc$ to be set ($\dM\!>\!10\GeVcc$, any $\thsbot$); leptons are
required in case of
$\ee\!\to\!\sto\bar{\sto}$, $\sto\!\to\!\b\ell\sneu$,
leading to a mass lower limit of $95\GeVcc$ ($\dM\!>\!10\GeVcc$, any $\thstop$). 
The stop decay $\sto\!\to\!\b\neu\f_{\u}\bar{\f}_{\d}$, recently
recognized as relevant, leads to a multi-body final state topology 
addressed by a dedicated ALEPH selection~\cite{ALEPH_squark_209}. As an
example, assuming the decay $\sto\!\to\!\b\chi\W^*$, the result is
$M_\sto\!>\!77\GeVcc$ ($\dM\!>\!10\GeVcc$, any $\thstop$).
ALEPH analyses also consider the case in which
a stop quasi-degenerate with the
LSP acquires a sizeable lifetime and hadronizes~\cite{ALEPH_stopldm}.
This scenario has been excluded searching for long-lived
heavy hadrons and an absolute stop mass lower limit of $63 \GeVcc$ has
been set for any $\thstop$, any branching ratio and
any $\dM$~\cite{ALEPH_squark_209}.

\subsection{Other topologies with $\emiss$ and $\pmiss$}
Topologies with two or more visible fermions in the final state plus
$\emiss$ and $\pmiss$ are expected in case of charginos and
neutralinos
production~\cite{DELPHI_ichep02,L3_ichep02,ALEPH_lsp_ichep02,OPAL_chaneu_ichep02}.
The processes are of the type 
$\ee\!\to\!\chi_{i>1}\chi$ and $\ee\!\to\!\chi_{i>1}\chi_{j>1}$ with
$\chi_{i>1}\!\to\!\chi\f\bar{\f}$, and $\ee\!\to\!\chi^{+}\chi^{-}$
with $\chi^\pm\!\to\!\chi\f_{\rm u}\f'_{\rm d}$. 
Cross section upper limits of $\sim\!0.1$--$0.3\pb$ are obtained by the
LEP-wide outcome of dedicated selections~\cite{LEPSUSYWG}.

Topologies with photon(s) can be very powerful in detecting new
phenomena~\cite{L3_ichep02,OPAL_ichep02,ALEPH_photons_209,DELPHI_photons_ichep02}.
Within the MSSM this case applies when heavier neutralinos 
are assumed to decay radiatively:
$\ee\!\to\!\chi_2\chi_2$ and $\ee\!\to\!\chi_2\chi$ with $\chi_2\!\to\!\chi\gamma$.
In this hypotheses the cross section upper limits range between $10\fb$ and 
$0.1\pb$ depending on the process~\cite{LEPSUSYWG}.

\section{INTERPRETATION}
\label{sec:int}
The negative results of the search for sparticle production
can be translated into constraints on the parameter space
in the context of specific SUSY models. Such a method allows the
exclusions to be extended to sparticles otherwise not accessible,
either because invisible, as the LSP, either because too heavy to be
produced~\cite{DELPHI_ichep02,L3_ichep02,OPAL_ichep02,ALEPH_lsp_ichep02}.

A widely accepted framework is the constrained MSSM (CMSSM). 
The unification of masses and 
couplings at the GUT scale allow the EW scale phenomenology to be set by
few parameters: $\tanb$, the ratio of the vacuum 
expectation values of the two Higgs doublets; $\mu$, the Higgs sector
mass parameter; $M_2$, the EW scale common gaugino    
mass; $m_0$, the GUT scale common scalar mass; the
trilinear couplings $A_\f$, that enter in the
prediction of the sfermion mixing and are generally set to fit the
no-mixing hypothesis.
\begin{figure}[t]
\includegraphics[width=0.43\textwidth]{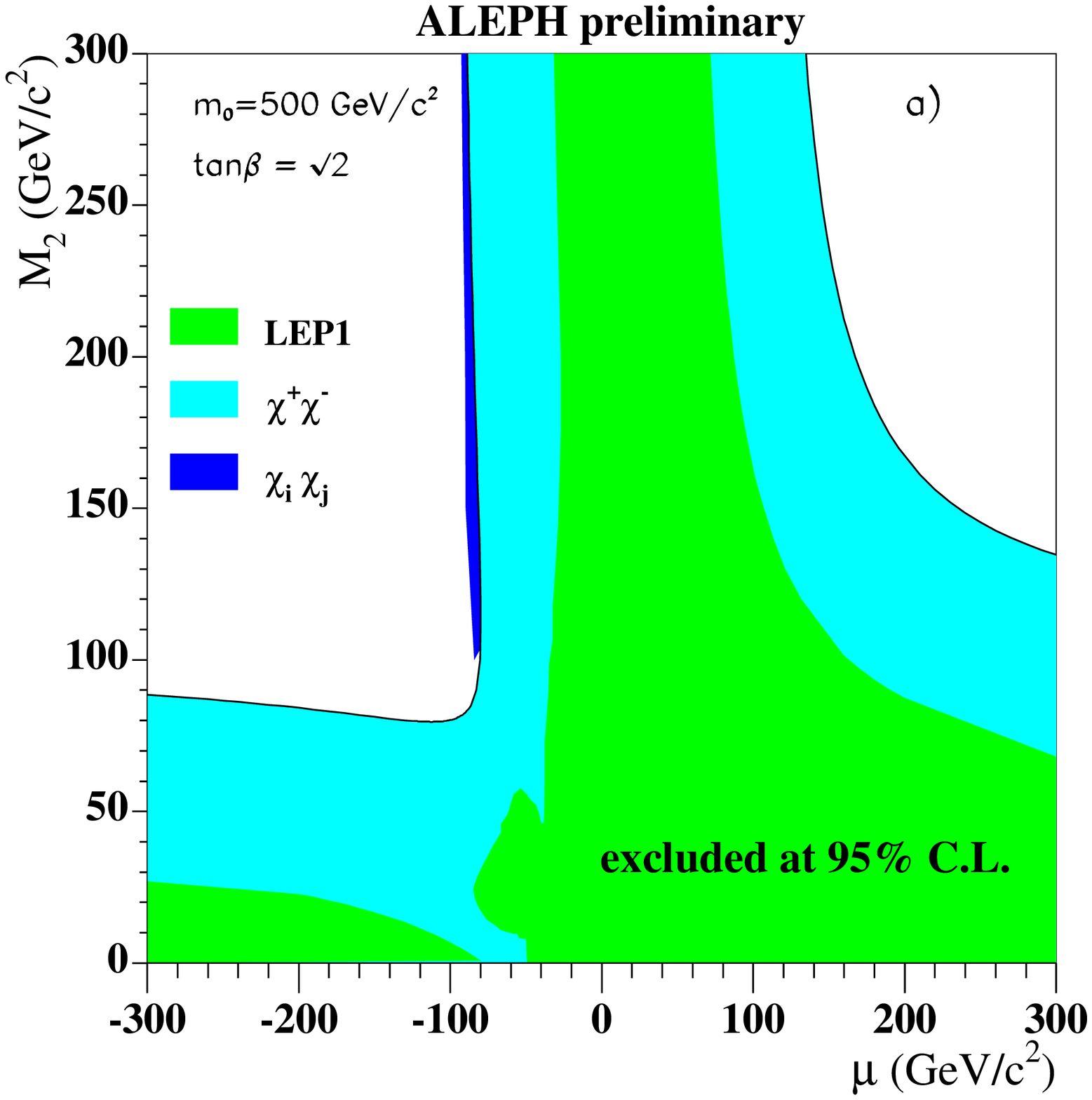} 
  \begin{picture}(0,0)
    \put(-98,103){
	\psframe[linecolor=white,fillstyle=solid,fillcolor=white](0,0)(2.8,2.9)
	} 
    \put(-106,152){
	\psframe[linecolor=white,fillstyle=solid,fillcolor=white](0,0)(0.5,1.18)
	} 
    \put(-46,98){
	\psframe[linecolor=white,fillstyle=solid,fillcolor=white](0,0)(1.05,0.5)
	} 
    \put(-106,98){
	\includegraphics[width=0.20\textwidth,clip]{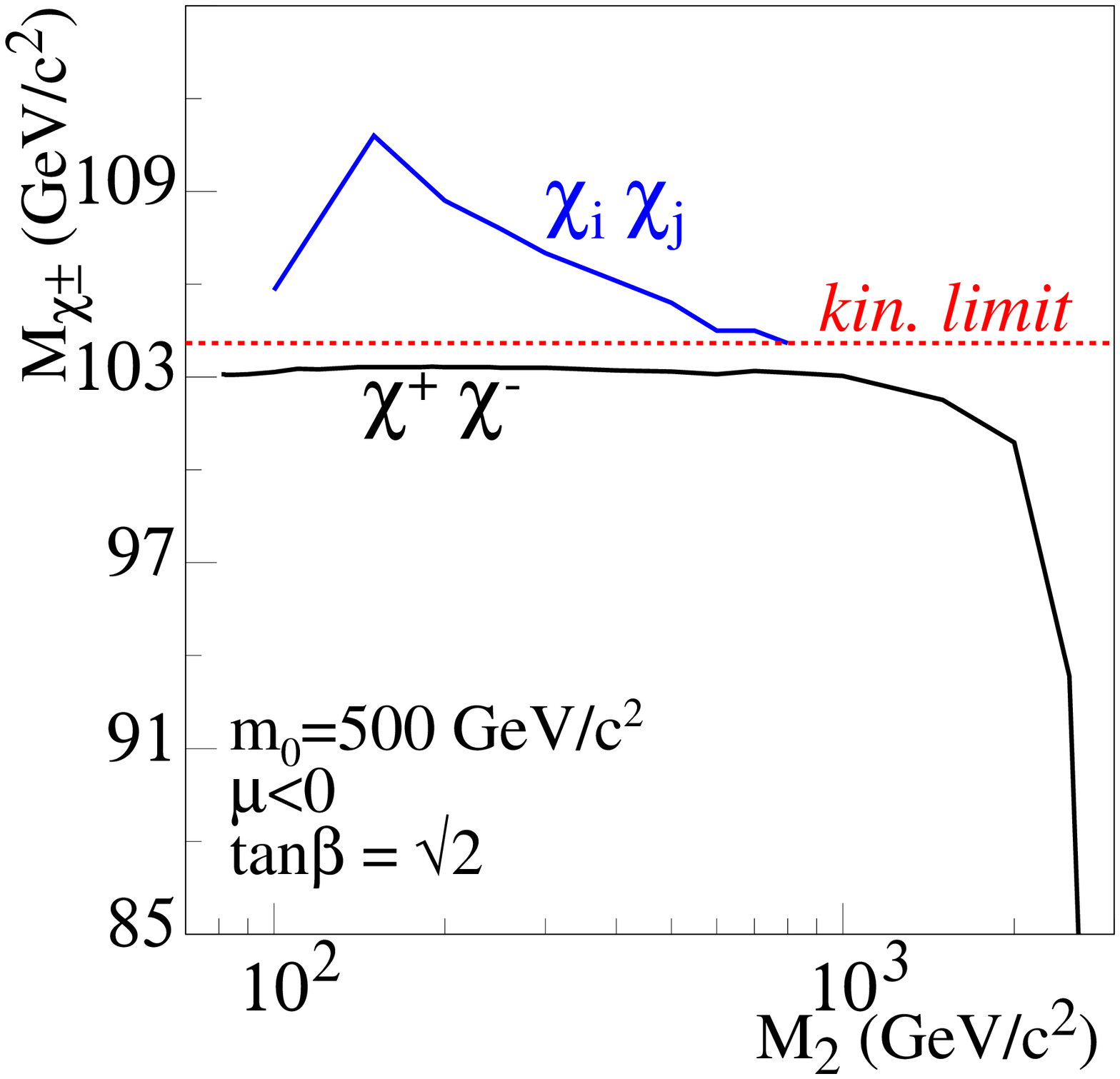}
	}
  \end{picture}
\vskip -1cm
\caption{Excluded domains in the $M_2$ vs. $\mu$ plane for
$\tanb\!=\!\sqrt{2}$ and $m_0\!=\!500\GeVcc$. The upper-right plot shows the
corresponding chargino mass lower limits for $\mu\!<\!0$.} 
\label{fig:mum2}
\end{figure}
\subsection{LSP limit}
\begin{figure}[t]
\includegraphics[width=0.355\textwidth]{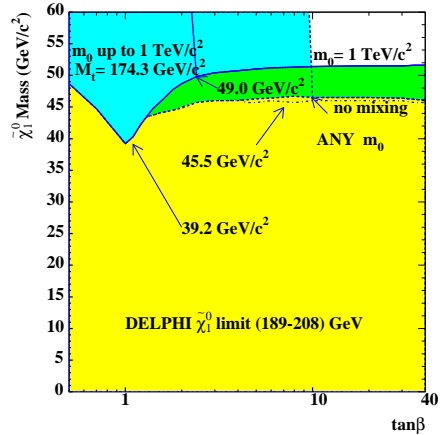} 
\vskip -1cm
\caption{The $\chi$-LSP limit vs. $\tanb$: large $m_0$
(solid curve) and any $m_0$ (dashed curve) in case of no-mixing;
any $m_0$ and $A_\tau\!=\!A_\b\!=\!A_\t\!=\!0$, i.e. mixing in the third family
(dash-dotted curve). The steep lines show the
impact of Higgs boson searches for two $m_0$ scenarios~\cite{DELPHI_ichep02}.} 
\label{fig:lsp}
\end{figure}

The negative outcome of charginos and neutralinos searches can be
used to exclude regions in the $(\mu,M_2)$ plane, as shown, as an
example, in Figure~\ref{fig:mum2} in which the sleptons are assumed
to decouple (i.e. large $m_0$). The upper-right plot of
Figure~\ref{fig:mum2} shows how neutralino searches allow chargino
exclusions to be improved for small $\tanb$ and $\mu\!<\!0$.
If the sleptons are lighter (small $m_0$ values), the chargino and
neutralino cross sections decrease for the enhancement
of negative-interfering slepton-exchange diagrams. The consequent
loss of sensitivity is recovered by slepton searches in such a
way that lower mass limits on gauginos and other sparticles as
$\seR$ or $\snu$ could be
set~\cite{DELPHI_ichep02,L3_ichep02,OPAL_ichep02,ALEPH_sel_ichep02}.
Among these, the most important is the LSP limit, i.e. the mass lower limit
on $\chi$, shown in Figure~\ref{fig:lsp} as a function 
of $\tanb$ for coupling and decoupling sleptons. The LSP mass lower limits
from LEP experiments fall between $36.3\GeVcc$ and
$39.6\GeVcc$ and are set for $\tanb\!\sim\!1$.

The LEP mass lower limits on the Higgs boson mass $m_\h$ can be also
used to further exclude small $\tanb$ ranges. Roughly, this 
just derives from the MSSM tree-level relation $m_\h\!<\!m_\Z|\!\cos
\beta|$. However, the details of the exclusion depend on $M_2$, $m_0$
and the stop mass because of the large radiative corrections to $m_\h$.
Adding the Higgs constraints the LSP mass lower limit substantially improves (up
to $\sim\!45\GeVcc$) and moves towards $\tanb\!\sim\!4$, as shown in
Figure~\ref{fig:lsp}.

\subsection{Impact of mixing in the third family}

The robustness of the LSP limit has been checked with respect to the
mixing effects in the third family, neglected in the above discussion. 
A stau getting light for mixing may be mass degenerate with the LSP,
making the chargino decays into staus difficult to
detect~\cite{DELPHI_ichep02,ALEPH_lsp_ichep02}.
Dedicated selections for
$\chi^\pm\!\to\!\st\nu_\tau\!\to\!\tau\chi\nu_\tau$ with soft taus,  
$\ee\!\to\!\chi_2\chi$ and $\ee\!\to\!\chi_2\chi_2$
with $\chi_2\!\to\!\tau\tau\chi$, and for chargino production in
association with an ISR photon ($\ee\!\to\!\chi^+\chi^-\gamma$) 
allow to solve this problem.
As shown in Figure~\ref{fig:lsp}, the LSP limits reported above have
been demonstrated to hold by using this studies, extended also considering
the mixing configurations for $\st$, $\sto$ and $\sbot$ that can
be explored by setting $A_\tau$, $A_\t$ and $A_\b$ to zero.

\subsection{mSUGRA}
\begin{figure}[t]
\includegraphics[width=0.44\textwidth]{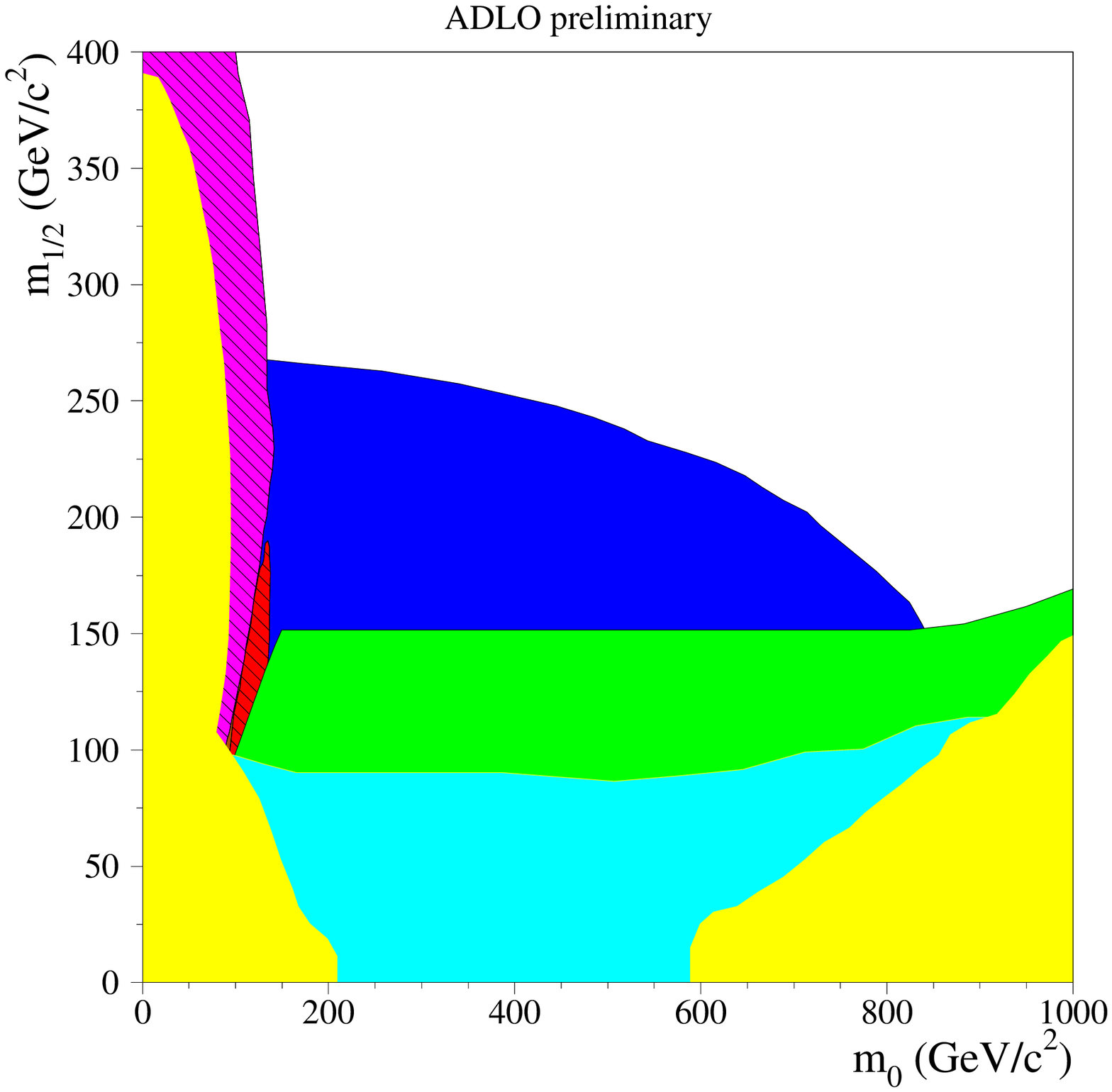} 
  \begin{picture}(0,0)
    \put(-112,91){
	\psframe[linecolor=white,fillstyle=solid,fillcolor=white](0,0)(2.95,3.00)
	} 
    \put(-112,92){
	\includegraphics[width=0.22\textwidth]{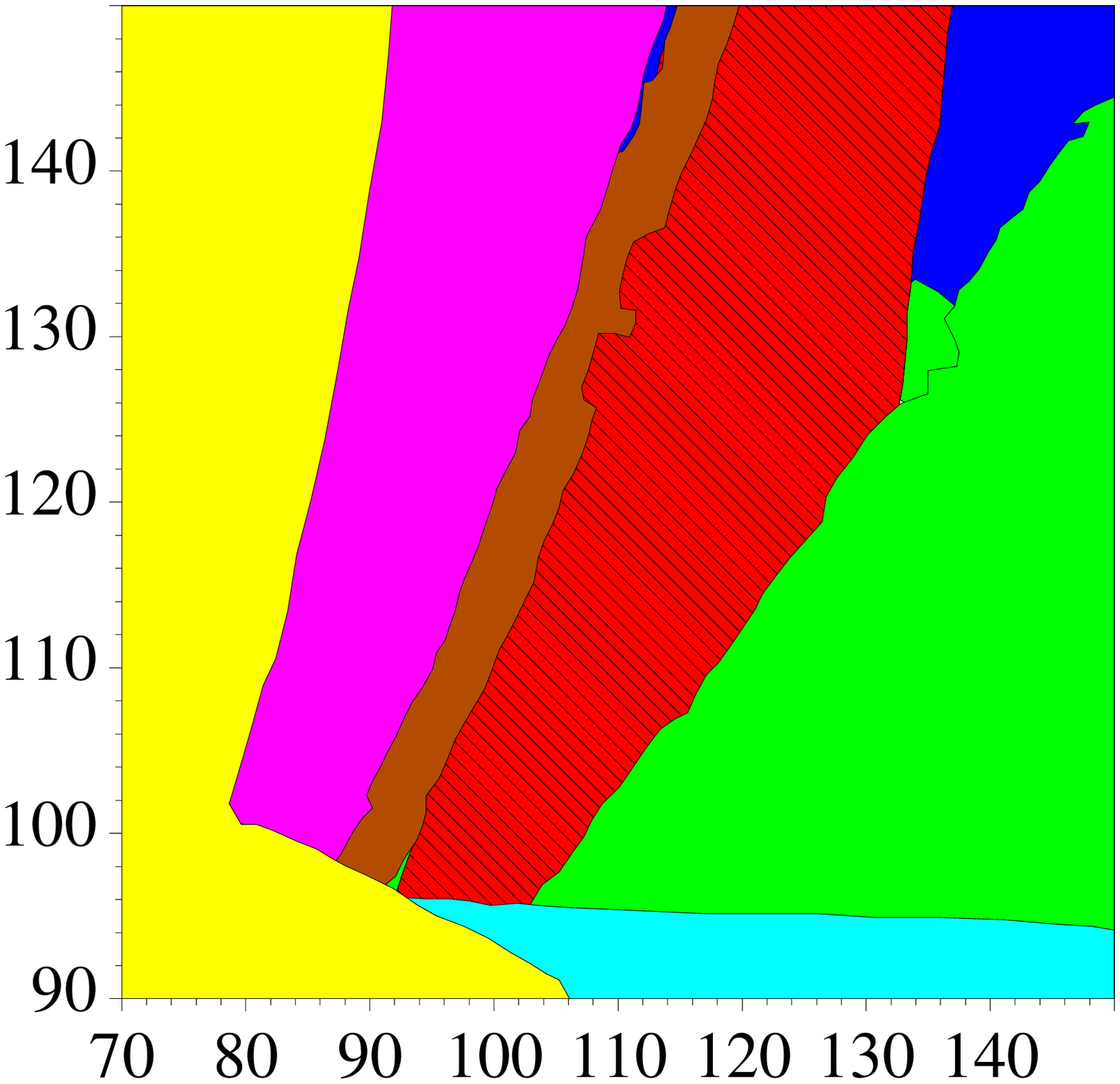}
	}
    \put(-50,35){\rotatebox{45}{\mbox{\footnotesize\bf Theory}}}
    \put(-47,25){\rotatebox{45}{\mbox{\footnotesize\bf forbidden}}}
    \put(-110,40){\mbox{\footnotesize\bf LEP1}}
    \put(-120,65){\mbox{\footnotesize\bf charginos}}
    \put(-145,90){\mbox{\footnotesize\white\bf Higgs}}
    \put(-66,114){\rotatebox{54}{\mbox{\footnotesize\white\bf sleptons}}}
    \put(-75,120){\rotatebox{67}{\mbox{\footnotesize\white\bf stau cascades}}}
    \put(-84,124){\rotatebox{74}{\mbox{\footnotesize\white\bf stable stau}}}
    \put(-177,38){\mbox{\footnotesize $\tanb\!=\!40$}}
    \put(-177,30){\mbox{\footnotesize $A_0\!=\!0$, $\mu\!<\!0$}}
    \put(-177,22){\mbox{\footnotesize $M_{\mathrm{top}}\!=\!175\mathrm{GeV\!/c^2}$}}
    \put(-169.5,56){
	\psframe[linecolor=white](0,0)(0.48,0.93)
	} 
    \put(-169.5,56){
	\psframe[linecolor=black,linestyle=dashed](0,0)(0.48,0.93)
	} 
    \put(-172,56){\rotatebox{90}{\mbox{\footnotesize zoom area}}}
  \end{picture}
\vskip -1cm
\caption{LEP combined exclusion domains in the mSUGRA $m_{1/2}$
vs. $m_0$ plane for $\tanb\!=\!40$, $A_0\!=\!0$ and $\mu\!<\!0$. A peculiar
area is zoomed to show the interplay between selections.}
\label{fig:mSUGRA}
\end{figure}
The results have been also interpreted within an even more constrained
version of the CMSSM, usually referred to as Minimal Supergravity (mSUGRA).
The relevant parameters are: $\tanb$, the sign of $\mu$ and $m_0$;
 $m_{1/2}$, the GUT scale common gaugino mass, that replaces
$M_2$; $A_0$, the GUT scale common trilinear coupling. 

On top
of LEP1 exclusions and theory-forbidden regions, small $m_0$ and $m_{1/2}$
areas are constrained from sleptons and gaugino searches, respectively. Higgs
boson searches are also effective, even in the large $\tanb$
range. As an example, Figure~\ref{fig:mSUGRA} illustrates $m_{1/2}$
vs. $m_0$ excluded domains for $\tanb\!=\!40$, $\mu\!<\!0$ and
$A_0\!=\!0$. The zoomed area focuses on the pathological region in which, for
the mixing, $\st$ and $\chi$ are almost degenerate and the
selections for stau-cascades and stable staus have to be used. 
The resulting mSUGRA LSP mass lower limits lie between $52$ and $59\GeVcc$,
depending on the top mass, and turn out to be $\sim\!8$--$9\GeVcc$ lower if $A_0$ is
allowed to assume values other than zero~\cite{LEPSUSYWG}.

\section{CONCLUSION}
Despite the negative outcome, LEP has substantially contributed to
the study of supersymmetric scenarios. Stringent constraints on
sparticle parameters have been set by direct search and by
interpretation studies within widely accepted frameworks. 
This huge experience is an important part of the LEP
legacy and it will result very useful for SUSY 
searches at future experiments.

\section*{ACKNOWLEDGMENTS}
I would like to thank G.~Ganis for the
careful reading of the manuscript. Special thanks
also to P.~Azzurri for his kind hospital-ity.

\end{document}